\documentclass[
  reprint,
 superscriptaddress,
bibnotes, 
aps,
pra,
]{revtex4-2}
\usepackage{float}
\usepackage[most]{tcolorbox}
\usepackage{enumitem}
\usepackage{amsfonts} 
\usepackage{booktabs}
\usepackage{graphicx}
\usepackage{dcolumn}
\usepackage{bm}
\usepackage{xcolor}
\usepackage{hyperref}      %
\usepackage{orcidlink}     
\usepackage{amsmath}
\usepackage{amssymb}
\usepackage{mathtools}
\usepackage[normalem]{ulem} 
\usepackage{xcolor}         
\usepackage{algpseudocode}

\usepackage{hyperref}

\begin{document}


\title{
Normalizing Flow-Based Bayesian Parameter Estimation for Noisy Quantum States}


\author{Hsien-Yi Hsieh\orcidlink{0000-0001-5227-8248}}
\affiliation{Institute of Photonics Technologies, National Tsing Hua University, Hsinchu 30013, Taiwan}

\author{Yuan-Ting Liu}
\affiliation{Institute of Photonics Technologies, National Tsing Hua University, Hsinchu 30013, Taiwan}

\author{Juan Camilo Rodr\'iguez\,\orcidlink{0009-0005-2128-3016}}
\affiliation{Department of Physics, National Tsing Hua University, Hsinchu 30013, Taiwan}

\author{Yang-Yi Lee}
\affiliation{Institute of Photonics Technologies, National Tsing Hua University, Hsinchu 30013, Taiwan}

\author{Po-Hang Wang\,\orcidlink{0009-0009-3800-5433}}
\affiliation{Institute of Photonics Technologies, National Tsing Hua University, Hsinchu 30013, Taiwan}

\author{Ole Steuernagel\orcidlink{0000-0001-6089-7022}}
\affiliation{Institute of Photonics Technologies, National Tsing Hua University, Hsinchu 30013, Taiwan}

\author{Chien-Ming Wu} 
\affiliation{Institute of Photonics Technologies, National Tsing Hua University, Hsinchu 30013, Taiwan}

\author{Ray-Kuang Lee\orcidlink{0000-0002-7171-7274}}
\email{rklee@ee.nthu.edu.tw}
\affiliation{Institute of Photonics Technologies, National Tsing Hua University, Hsinchu 30013, Taiwan}
\affiliation{Department of Physics, National Tsing Hua University, Hsinchu 30013, Taiwan}
\affiliation{Center for Theory and Computation, National Tsing Hua University, Hsinchu 30013, Taiwan}
\affiliation{Center for Quantum Science and Technology, Hsinchu 30013, Taiwan}

\date{\today}

\begin{abstract}

To extract complete information about experimental noisy quantum states, as well as correlations among different physical parameters, we develop an efficient machine learning–assisted framework with the normalizing flow based Bayesian parameter estimation (BPE).  
By embedding a physically interpretable parameter vector, $\boldsymbol{s}$, into the quantum state ansatzes, $\rho(\boldsymbol{s})$,  the  corresponding parameter distribution~$p(\boldsymbol{s}|D)$ is induced from BPE, based on the available experimental data $D$.
The BPE allows us to perform parameter correlation analyses, providing powerful insights about how experimental parameters and related noise sources affect quantum features of the system. 
As an example, the flexibility of our framework is demonstrated using two different physical ansatzes on the optical cat state experiments, where a noisy single-photon state is added to impure squeezed state.
With correlations among the  Wigner function negativity, photon-addition fraction, noisy fraction, and pump powers, our framework gives interpretations for complicated experimental mechanisms in generating noisy non-Gaussian states.
By leveraging normalizing flow-based neural networks and Bayesian uncertainty estimation, along with physical and experimental constraints incorporated naturally,  the valuable information inferred in the posterior learning provides guidelines for experimentalists, as well as theorists, in identifying promising parameters to enhance desirable features. 
\end{abstract}
\maketitle


\noindent {\it Introduction.}~~Quantum state reconstruction has so far focused primarily on determining the best-fit state $\rho$, i.e., the state that, based on the available experimental data $D$, `best' conforms with $D$. Within this point estimation paradigm, one widely adopted method for continuous\textcolor{blue}{-}variable quantum states is maximum-likelihood estimation (MLE), which has been extensively studied and refined over the past two decades~\cite{RevModPhys}, including iterative MLE (iMLE)~\cite{Lvovsky_2004} and, more recently, machine learning-assisted MLE approaches~\cite{tiunov2020experimental,Lvosky_NN_QST}. An alternative approach is Bayesian estimation~\cite{Lu2022_BPEQST,Chapman22_BPEQST}, which infers a posterior distribution over density matrices by combining the experimental data with an appropriate prior distribution. A connection to the point estimation can be established through the posterior mean, $\hat{\rho}=\int \rho ~p(\rho|D)~d\rho$~\cite{Hradil_book_theory}.

However, in practical experiments we care more about interpretable parameters in generating quantum states than the resulting density matrices. If a quantum state can be parameterized by a set of physically meaningful parameters $\boldsymbol{s}$, Bayesian inference can instead be performed in this comparatively simpler parameter space, yielding the posterior $p(\boldsymbol{s}\mid D)$, a classical probability distribution over $\boldsymbol{s}$. Representative point estimates may then be obtained from appropriate posterior summaries, such as the posterior mean, $\hat{\boldsymbol{s}}=\int \boldsymbol{s}~p(\boldsymbol{s}| D)~d\boldsymbol{s}$, or the marginal posterior median~\cite{berger1985statistical,kruschke2015doing}. 
In addition, the parameters  $\boldsymbol{s}$ can directly describe physical mechanisms, including settings of classical (experimental) parameters and their associated noise sources. We emphasize that we do not merely strive to determine a single point estimate, as such a summary ignores valuable information encoded in $p(\boldsymbol{s}| D)$. It is the distribution $p(\boldsymbol{s}|D)$ we infer, as it provides a more complete picture of the inferred parameters, as well as their uncertainties and mutual dependencies.

In this {\it Letter}, we propose a Bayesian framework for inferring the parameters $\boldsymbol{s}$ for physically motivated quantum state ansatzes $\rho(\boldsymbol{s})$,  directly from homodyne measurement data. To perform inference efficiently in experimentally relevant settings, we employ normalizing flow-based variational inference~\cite{flow_review} to approximate the posterior distribution $p(\boldsymbol{s}|D)$. Since an inappropriate ansatz may lead to biased physical interpretations, we further assess different physically motivated ansatzes against the experimental data and analyze the corresponding  parameter uncertainties, correlations, and the underlying experimental processes.
We demonstrate this flexibility for optical cat states generated by photon-addition, where several intricate physical processes may contribute to the observed data~\cite{photon_add_exp}. 
Compared with full quantum state reconstruction, our approach provides direct insight into how experimental parameters and related noise sources affect the non-Gaussian state preparation and degradation, offering a systematic route to identify desirable and unwanted physical processes in real world quantum optics experiments.


\medskip

\noindent {\it Variational inference on parametrized quantum states.}~~
Unlike the iMLE homodyne tomography, where the quantum state $\rho$ is reconstructed directly from measurement data~\cite{Lvovsky_2004}, our inference is performed over the physical parameter space $\boldsymbol{s}$ underlying $\rho(\boldsymbol{s})$. For example, a squeezed vacuum state
$
\rho_{\mathrm{sq}}
=
\hat{S}(r,\phi)\lvert0\rangle\langle0\rvert\hat{S}^{\dagger}(r,\phi)
$
is fully specified by $\boldsymbol{s}=(r,\phi)$, with the squeezing strength $r$ and squeezing angle $\phi$.
For a noisy single-photon state with support truncated at the two-photon Fock level, we have
$\rho=c_0\lvert0\rangle\langle0\rvert+c_1\lvert1\rangle\langle1\rvert+c_2\lvert2\rangle\langle2\rvert$~\cite{Fock_QST}.
Here, the normalization constraint $c_0+c_1+c_2=1$ reduces the independent parameter set to $\boldsymbol{s}=(c_0,c_1)$. For optical cat state generated by photon-addition, via stimulated parametric down conversion (StPDC)~\cite{photon_add_exp}, we employ a physically motivated ansatz parametrized by $
\boldsymbol{s}
=
(p_1,p_2,p_3,\eta,r,\phi,\bar n_1,\bar n_2),
$ with
\begin{equation}
\rho
=
\eta\,
\frac{a^\dagger\rho_{\mathrm{deg}}a}
{\operatorname{Tr}[a^\dagger\rho_{\mathrm{deg}}a]}
+
(1-\eta)\, \rho_{\mathrm{deg}},
\label{eq:spdc_model}
\end{equation}
where the degraded component follows the experimentally motivated mixture
model of Ref.~\cite{SQ_PRL,Rodriguez:25,SQ_symmetry}
\begin{equation}
\rho_{\mathrm{deg}}
=
p_1\, \rho_{\mathrm{sq}}(r,\phi)
+
p_2\, \rho^\mathrm{sq}_\mathrm{th}(r,\phi,\bar n_1)
+
p_3\,\rho_{\mathrm{th}}(\bar n_2)
\label{eq:sq_model}
\end{equation}
with $p_1+p_2+p_3=1$. Here, $\eta\in[0,1]$ denotes the photon-added fraction, while $\bar n_1$ and $\bar n_2$ characterize the thermal contributions. Thus, the state is fully specified by a low-dimensional, physically interpretable parameter vector $\boldsymbol{s}$, which forms the target of our Bayesian inference.

\begin{figure}[t]
    \centering
    \includegraphics[width=\columnwidth]{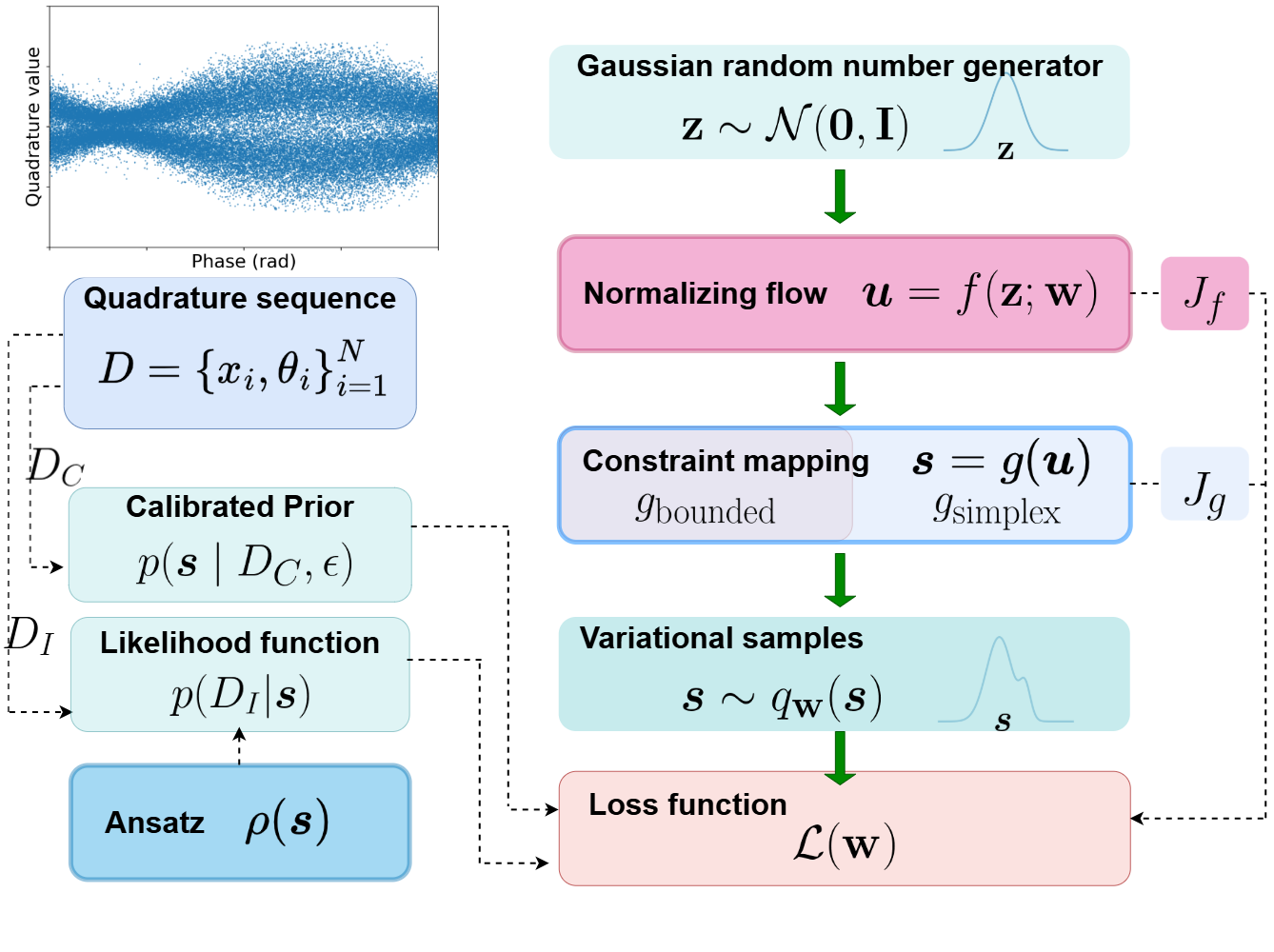}
    \caption{Schematic illustration of the forward-pass architecture of the
normalizing flow-based inference model.
The experimental data are split into calibration ($D_C$) and inference ($D_I$) subsets. The calibration data $D_C$, together with empirical knowledge $\epsilon$, are used to construct the calibrated prior, while $D_I$ enters the likelihood defined by the chosen physical ansatz $\rho(\mathbf{s})$. A single forward pass (green arrows) transforms Gaussian latent samples through the normalizing flow and subsequent constraint mappings, yielding samples from the flexible variational distribution
$q_{\mathbf{w}}(\mathbf{s})$ that satisfy the prescribed physical and
empirical constraints.
The two constraint mappings respectively impose bounded and simplex
parameterizations.
The inset distributions schematically show one-dimensional projections
of the latent and transformed distributions.
The loss incorporates the likelihood function, the calibrated prior,
and the corresponding Jacobian contributions $J_f$ and $J_g$;
optimization of $\mathbf{w}$ then drives
$q_{\mathbf{w}}(\mathbf{s})$ toward the target posterior.}
    \label{fig:architecture}
\end{figure}

The forward-pass architecture of our normalizing flow-based BPE is schematically illustrated in Fig.~\ref{fig:architecture}. Following an empirical Bayes-inspired strategy~\cite{Casella1985}, we partition each homodyne dataset $D$ into a calibration subset $D_C$ (20\%) and an inference subset $D_I$ (80\%). The calibration data, together with empirical prior information $\epsilon$, are used to construct the prior $p(\boldsymbol{s}|D_C,\epsilon)$, whereas $D_I$ enters exclusively through the likelihood. Here, $\epsilon$ collectively represents empirical prior information used
to specify physically plausible parameter ranges and candidate parameter configurations, based on physical considerations, experimental experience, or auxiliary experimental characterization. This information is distinguished from $D_C$, which is used to select a coarse representative configuration from these candidates and thereby determine the prior means. Further details are provided in \ref{sec:prior calibration} and \ref{sec:sa_measurements} of the Supplementary Material. Assuming that, conditioned on $\boldsymbol{s}$, the inference data $D_I$ are independent of the calibration information $(D_C,\epsilon)$, the resulting posterior takes the form
\begin{equation}
p(\boldsymbol{s}| D_I,D_C,\epsilon)
\propto
p(D_I|\boldsymbol{s})\,
p(\boldsymbol{s}| D_C,\epsilon).
\end{equation}
This separation prevents the same measurements from being used, both to specify the prior and to perform posterior inference.

We approximate the posterior distribution $p(\boldsymbol{s}| D_I,D_C,\epsilon)$ using a variational distribution $q_s(\boldsymbol{s};\mathbf{w})$, where $\mathbf{w}$ denotes the neural network parameters of the normalizing flow, and optimize $\mathbf{w}$ by minimizing the Kullback--Leibler divergence (KL, denoted as $D_{\mathrm{KL}}$)~\cite{KL_divergence}:
\begin{equation}
\mathcal{L}(\mathbf{w})= \mathbb{E}_{q_s(\boldsymbol{s};\mathbf{w})}
\left[
\log q_s(\boldsymbol{s};\mathbf{w})
- \log p(\boldsymbol{s}| D_I,D_C,\epsilon)
\right].
\label{Loss fun_1}
\end{equation}
Applying Bayes' theorem and absorbing terms independent of $\mathbf{w}$ into a constant yields $\mathcal{L}(\mathbf{w})
=
\mathbb{E}_{q_s(\boldsymbol{s};\mathbf{w})}
\left[
\log q_s(\boldsymbol{s};\mathbf{w})
-
\log p(D_I|\boldsymbol{s})
-
\log p(\boldsymbol{s}| D_C,\epsilon)
\right]
+
C$, where $C$ is a constant independent of the variational parameters $\mathbf{w}$, and $p(D_I\mid\boldsymbol{s})$ is the likelihood of the inference data.

We represent the variational distribution $q_s(\boldsymbol{s};\mathbf{w})$ through a flow-based transformation that maps samples from a simple base distribution to physically admissible parameter values. Specifically, we define
$\boldsymbol{s}=T(\mathbf{z};\mathbf{w})
=
g\left(f(\mathbf{z};\mathbf{w})\right)$,
where $\mathbf{z}\sim p_z(\mathbf{z})$, $f$ is a normalizing flow~\cite{flow_review,PhysRevLett.130.171402,PhysRevD.109.123547,prb_flow,PhysRevLett.127.062701}, implemented here as a neural spline flow (NSF)~\cite{spline_flow}. And $g$ is a deterministic constraint mapping enforcing simplex and bounded domain constraints on physical parameters (Details of $g$ are provided in \ref{sec:constraint mapping} of Supplementary Materials). The constraint mapping is designed such that samples are restricted to the support of corresponding prior distributions. The variational posterior is obtained through the change of variable formula: 
$q_s(\boldsymbol{s};\mathbf{w})
=p_z(\mathbf{z})
\left|
\det J_T(\mathbf{z};\mathbf{w})
\right|^{-1}
$, and $\det(J_T)=\det(J_g)\det(J_f)$.

Since $T=g\circ f$, the Jacobian factorizes into contributions from the flow and constraint mappings. The Jacobian associated with $g$ admits a closed-form expression and can therefore be computed without numerical approximation. Substituting the above relation into the KL objective yields 
\begin{equation}
\begin{aligned}
\mathcal{L}(\mathbf{w})
={}&
\mathbb{E}_{p_z(\mathbf{z})}
\Big[
\log p_z(\mathbf{z})
-\log \left|\det J_T(\mathbf{z};\mathbf{w})\right|
\\
&\quad
-\log p(D_I|\boldsymbol{s})
-\log p(\boldsymbol{s}|D_C,\epsilon)
\Big]+C, 
\end{aligned}
\label{eq:MC estimate}
\end{equation}
which is optimized using Monte-Carlo samples from the base distribution $p_z$. 
Fig.~\ref{fig:architecture} summarizes the corresponding variational inference workflow.

The training samples $\mathbf{z}\in \mathbb{R}^{n-1}$ drawn from a simple base distribution (Gaussian) are transformed by a neural network $\boldsymbol{u}=f(\mathbf{z};\mathbf{w})$, with $\boldsymbol{u}\in \mathbb{R}^{n-1}$, into samples of a more complex target distribution. A subsequent deterministic mapping $g$ transforms part of the output into parameters $\boldsymbol{s}_{\mathrm{simplex}}\in\Delta^{m-1}$ constrained to sum to unity, while the remaining outputs are mapped to bounded parameters $\boldsymbol{s}_{\mathrm{bounded}}\in\prod_{j=1}^{n-m}[l_j,h_j]$, where $l_j$ and $h_j$ denote the lower and upper bounds of the $j$th bounded parameter, respectively. The constraint mapping $g$ is constructed such that its image coincides with the support of the corresponding prior distributions, thereby ensuring physically admissible parameter transformations throughout training. 
The full set of inferred physical parameters is therefore
$\boldsymbol{s}=(\boldsymbol{s}_{\mathrm{simplex}},
\boldsymbol{s}_{\mathrm{bounded}})$, containing $n$ components but
$n-1$ independent degrees of freedom due to the simplex constraint, as exemplified by $p_1+p_2+p_3=1$ in Eq.~(\ref{eq:sq_model}).
Here, $n$ is determined by the number of parameters in the chosen
physical ansatz $\rho(\boldsymbol{s})$. Accordingly, the architecture can be readily adapted to different parameterized quantum-state models by adjusting its dimensionality and constraint mapping to the corresponding ansatz.

For the likelihood construction, each record $(x_i,\theta_i)\in D_I$ is treated as an independent measurement conditioned on the local oscillator phase $\theta_i$. The corresponding quadrature probability density is given by
$p_{\theta_i}(x_i|\boldsymbol{s})=
\operatorname{Tr}\left[
\Pi(\theta_i,x_i)\rho(\boldsymbol{s})
\right]$,
where $\Pi(\theta_i,x_i)$ denotes the POVM element associated with homodyne detection at phase $\theta_i$~\cite{Lvovsky_2004}. Assuming independent measurements, the likelihood factorizes over the inference dataset, yielding
$\log p(D_I|\boldsymbol{s})=\sum_{(x_i,\theta_i)\in D_I}\log p_{\theta_i}(x_i|\boldsymbol{s})$.
The log-likelihood is evaluated numerically using the density matrix represented in the Fock basis, with the computation vectorized and executed on GPU hardware using {\it PyTorch}~\cite{PyTorch}. The likelihood evaluation itself is identical to that used in the standard iMLE framework.

The calibrated prior $p(\boldsymbol{s}\mid D_C,\epsilon)$ is constructed
using a Dirichlet distribution over simplex-constrained parameters and
independent Beta distributions over bounded physical parameters~\cite{Piech2023Probability,Lin2016Dirichlet}. For the photon-addition model of Eq.~(\ref{eq:spdc_model}), these correspond to
$\boldsymbol{s}_{\mathrm{simplex}}=(p_1,p_2,p_3)$ and
$\boldsymbol{s}_{\mathrm{bounded}}
=(\eta,r,\phi,\bar n_1,\bar n_2)$,
respectively. The resulting priors are chosen to remain weakly informative. Details of their parameterization and calibration are provided in
Supplementary Material~\ref{sec:prior calibration}
and~\ref{sec:sa_measurements}.

\medskip
\noindent {\it Analysis of experimental data.}~~
As an experimental demonstration, we apply our general inference framework
to non-Gaussian optical cat states. Originating from Schr\"odinger's
gedanken experiment~\cite{Schrodinger1935}, cat states have been realized
across a variety of physical platforms, including quantum optics~\cite{Ourjoumtsev2006,NeergaardNielsen2006,
ourjoumtsev2007generation,ourjoumtsev2009preparation,
sychev2017enlargement}, trapped ions~\cite{cat-ion}, superconducting
qubits~\cite{vlastakis2013deterministically}, and optomechanical
systems~\cite{OM-cat}. Their non-classical phase space features
~\cite{Braunstein2005,Weedbrook2012} make them relevant both for probing
the quantum-to-classical boundary and for quantum information processing,
including cat-code architectures~\cite{chamberland2022building,
gravina2023critical} and the preparation of quantum error-correcting
codes~\cite{gottesman2001encoding,vasconcelos2010all,eaton2019nongaussian,
tzitrin2020progress,bourassa2021blueprint}.

In optical implementations, cat states can be engineered by photon-subtraction or photon-addition on squeezed states~\cite{zavatta2004quantum}.
Photon-subtraction is commonly implemented through conditional
measurements at a beam splitter~\cite{Wakui:07,takahashi2008generation,
asavanant2017generation,takase2021generation}, whereas photon-addition can
be heralded by injecting a squeezed state into a parametric down-conversion
process~\cite{arman2021photon,photon_add_exp}. The coexistence of imperfect
squeezing, thermal degradation, and finite heralding efficiency makes these
non-Gaussian states a particularly useful case study for our framework:
rather than merely estimating a state, the inferred posterior can resolve
the contributions and uncertainties associated with multiple physical
mechanisms.

We therefore analyze two sets of homodyne data obtained from heralded
photon-added squeezed states~\cite{photon_add_exp}. 
In experiments, the Optical Parametric Oscillator (OPO) controls the preparation of the squeezed input state, whereas the StPDC process provides the single-photon state to be added. 
In {\it Dataset 1},
the OPO pump power is varied while the StPDC setting is fixed.
On the other hand, in {\it Dataset 2}, the StPDC pump power is varied while the OPO setting is fixed.
These two datasets therefore allow us to examine how the inferred physical mechanisms evolve under distinct experimental controls.

Our objective is to investigate the extent to which physically meaningful
mechanisms can be inferred from each homodyne dataset through a single-shot
Bayesian analysis. Throughout this work, posterior distributions are inferred
independently for individual datasets, without iterative experimental updates,
allowing us to focus on the intrinsic inference capability of the proposed
framework.

The homodyne quadratures are calibrated using the vacuum state and normalized
such that
$\mathrm{Var}(\hat X_\theta)=1/4$.
In addition, three independent squeezing/anti-squeezing characterization
measurements are used to define physically motivated prior supports for the
squeezing parameters; the corresponding measurements are described in ~\ref{sec:sa_measurements} of Supplementary Materials.

To demonstrate the flexibility of the inference framework, we also consider two
physically motivated ansatzes. The first ansatz $\mathcal A_1$ is the photon-addition
model defined in Eq.~(\ref{eq:spdc_model}). More than that, we account the possible
displacement induced by back-scattered light into account, by introducing the second ansatz $\mathcal A_2$,
\begin{eqnarray}
\rho
=\, && \eta\,  D(\beta_1)
\left(
\frac{a^\dagger\rho_{\mathrm{deg}}a}
{\mathrm{Tr}[a^\dagger\rho_{\mathrm{deg}}a]}
\right)
D^\dagger(\beta_1)
\nonumber\\
&& +(1-\eta)\,
D(\beta_0)\rho_{\mathrm{deg}}D^\dagger(\beta_0),
\label{eq:back}
\end{eqnarray}
which introduces four additional parameters for the displacement
amplitudes ($|\beta_{0,1}|$) and phases ($\arg(\beta_{0,1})$), where
the subscripts $0$ and $1$ denote the non-added and single-photon-added
branches, respectively.
The resulting 12-parameter inference problem is handled by the same
variational workflow without modification.

For each experimental setting, Bayesian inference is performed independently
for the two ansatzes, labeled by $a=1,2$. A representative parameter
vector $\hat{\boldsymbol{s}}_a$ is formed from the posterior mean of the
simplex weights $(p_1,p_2,p_3)$, preserving their normalization, and the
posterior medians of the remaining parameters. We then construct
$\hat{\rho}_a=\rho(\hat{\boldsymbol{s}}_a)$ and compare its log-likelihood
with the iMLE reconstruction through

\begin{equation}
\Delta\ell_a
=
\frac{1}{N}
\left[
\log p(D_I|\hat{\rho}_a)
-
\log p(D_I|\hat{\rho}_{\mathrm{iMLE}})
\right],
\label{eq:predictive_ll}
\end{equation}
where $N$ denotes the number of quadrature samples used in the comparison.
The ansatz with the larger log-likelihood is taken as the better-supported
description for the corresponding experimental setting.

\begin{figure}[t]
    \centering
    \includegraphics[width=8.4cm]{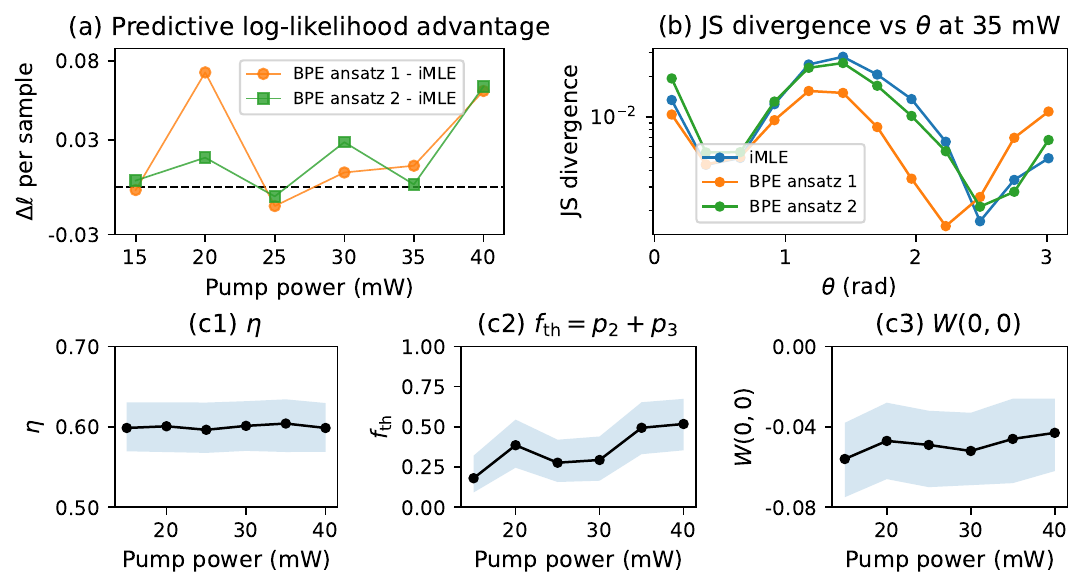}
    \caption{Predictive performance and inferred physical parameters for
    {\it Dataset 1}, with two BPE ansatzes.
    (a) Average log-likelihood difference, $\Delta\ell=\ell_{\rm BPE}-\ell_{\rm iMLE}$, per quadrature sample, with the dashed line marking $\Delta\ell=0$; positive values indicate improved agreement relative to the iMLE reconstruction.
    (b) Phase-resolved Jensen--Shannon (JS) divergence between the measured
    and predicted quadrature distributions for the 35 mW OPO setting.
    Posterior medians of (c1) the photon-addition fraction $\eta$,
    (c2) the thermal fraction $f_{\rm th}=p_2+p_3$ defined in Eq.~(\ref{eq:sq_model}), and
    (c3) the Wigner function at the origin $W(0,0)$ as functions of OPO pump
    power. Shaded regions denote $68\%$ credible intervals.}
    \label{fig:main_opo}
\end{figure}

Fig.~\ref{fig:main_opo}(a) summarizes the likelihood comparison for
{\it Dataset 1}, with two BPE ansatzes. Across most OPO pump powers, at least one of the two physical ansatzes achieves a likelihood comparable to or greater than that
of the iMLE reconstruction, indicating that the parametrized models do capture
the dominant features of the measured quadrature statistics. The 25 mW dataset constitutes the main exception, for which both models
show a small negative average log-likelihood difference,
$\Delta\ell=\ell_{\rm BPE}-\ell_{\rm iMLE}$, suggesting that additional experimental effects may not be fully represented by either ansatz. This may reflect additional experimental noise in this dataset, such as phase fluctuations or other measurement-induced imperfections.

As a complementary comparison,
Fig.~\ref{fig:main_opo}(b) shows the phase-resolved
Jensen--Shannon (JS) divergence~\cite{JS_divergence}
for the 35 mW dataset, which provides a direct view of how the model--data agreement varies with the local-oscillator phase. Here, for each local-oscillator phase bin, we evaluate
\begin{equation}
\begin{aligned}
\mathrm{JS}(\theta)
={}&
\frac{1}{2}D_{\mathrm{KL}}(p_\theta\Vert m_\theta)
+
\frac{1}{2}D_{\mathrm{KL}}(q_\theta\Vert m_\theta),
\\[-2pt]
&m_\theta=\frac{1}{2}(p_\theta+q_\theta),
\end{aligned}
\label{eq:JS}
\end{equation}
where $p_\theta$ and $q_\theta$ denote the empirical and model-predicted
quadrature distributions, respectively, with smaller values indicating
closer agreement. Consistent with Fig.~\ref{fig:main_opo}(a), ansatz $\mathcal A_1$ generally
yields smaller JS divergences over most phases, although this tendency
reverses near $\theta=\pi$, showing that the two measures emphasize
different aspects of the measured quadrature statistics.

Posterior samples provide uncertainty estimates not only for the inferred
parameters but also for derived state quantities, such as $W(0,0)$. Throughout the following,
point estimates are represented by posterior medians, with shaded regions
denoting the corresponding $68\%$ credible intervals. As shown in Figs.~\ref{fig:main_opo}(c1)--(c3), the photon-addition fraction $\eta$ remains comparatively stable as the OPO pump power is increased, whereas the thermal fraction $f_{\rm th}=p_2+p_3$ increases, while $W(0,0)$ becomes less negative.

\begin{figure}[t]
    \centering
    \includegraphics[width=8.4cm]
    {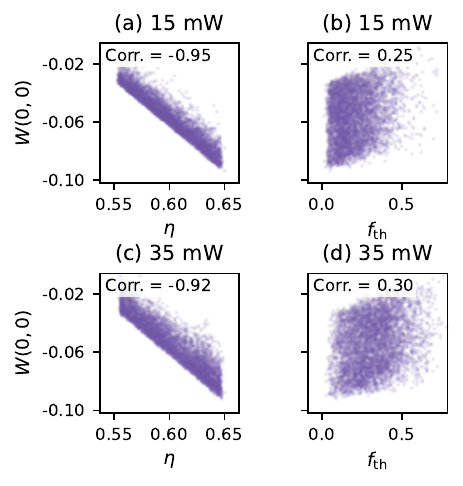}
    \caption{
    Posterior correlations between the inferred photon-addition fraction
    $\eta$ or thermal fraction $f_{\rm th}=p_2+p_3$ and the Wigner function
    at the origin $W(0,0)$ for {\it Dataset 1}.
    Panels (a) and (b) correspond to the 15 mW OPO setting, while
    panels (c) and (d) correspond to the 35 mW setting.
    Corr. denotes the Pearson correlation coefficient evaluated from
    posterior samples.
    }
    \label{fig:corr_opo}
\end{figure}

Beyond the marginal trends, the joint posterior distributions reveal how
different inferred mechanisms are related to the resulting state
nonclassicality. Fig.~\ref{fig:corr_opo} shows representative posterior
samples for the 15 and 35 mW OPO settings. A strong anticorrelation is
observed between the photon-addition fraction $\eta$ and Wigner negativity $W(0,0)$,
indicating that larger photon-addition contributions are associated with
more negative values of the Wigner function. In contrast, the correlation
between the thermal fraction $f_{\rm th}$ and Wigner negativity $W(0,0)$ is substantially
weaker. Importantly, these relationships emerge from the joint posterior
inferred from the homodyne measurements rather than being imposed as
explicit constraints on the model.

An analogous analysis is performed for {\it Dataset 2}, in which the StPDC
pump power is varied while the OPO setting is fixed. The corresponding
likelihood comparison, phase-resolved JS divergence, and marginal posterior
trends are provided in~\ref{sec:dataset2} of Supplementary Materials.
Here, we focus on comparing the posterior correlation structure with that
obtained for {\it Dataset 1}.

\begin{figure}[t]
    \centering
    \includegraphics[width=8.4cm]
    {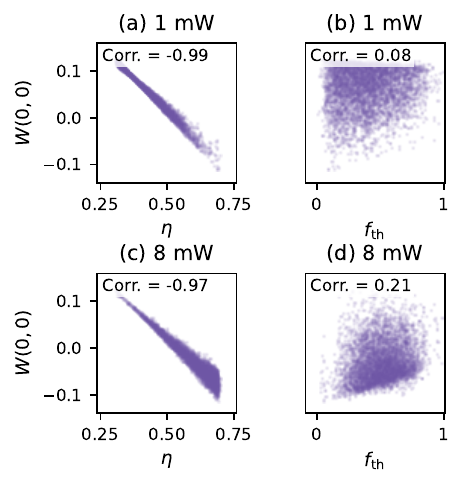}
    \caption{
    Posterior correlations between the inferred photon-addition fraction
    $\eta$ or thermal fraction $f_{\rm th}=p_2+p_3$ and the Wigner function
    at the origin $W(0,0)$ for {\it Dataset 2}.
    Panels (a) and (b) correspond to the 1 mW StPDC setting, while
    panels (c) and (d) correspond to the 8 mW setting.
    Corr. denotes the Pearson correlation coefficient evaluated from
    posterior samples.
    }
    \label{fig:corr_spdc}
\end{figure}

As shown in Fig.~\ref{fig:corr_spdc}, the strong anticorrelation between
the photon-addition fraction $\eta$ and the Wigner function at the origin
$W(0,0)$ persists when the StPDC pump power is varied, with Pearson
coefficients~\cite{Pearson1895} close to $-1$ for both representative settings. By contrast, the correlation between the thermal fraction $f_{\rm th}$ and $W(0,0)$ remains weak and is even less pronounced than in {\it Dataset 1}. The corresponding pump power dependence in the \ref{sec:dataset2} of the Supplementary Material also gives a more nonmonotonic trend than that observed for the OPO series, indicating that the evolution of the Wigner negativity cannot be attributed to a single degradation parameter. Instead, it reflects the joint evolution of photon addition, thermal mixing, and the remaining physical parameters.

\medskip

\noindent {\it Conclusion.}~~We have developed a Bayesian inference framework that naturally incorporates physical and experimental constraints through both parameterized quantum state models and prior distributions. 
By combining variational inference with normalizing flows, the proposed approach enables efficient posterior inference while respecting physically admissible parameter domains. Our approach naturally provides both point estimates and uncertainty quantification while enabling systematic investigation of competing physical hypotheses. 

As an illustration, we apply our normalizing flow-based Bayesian parameter estimation to optical cat states with photon-addition, with varying OPO and StPDC pump powers.
For the non-Gaussian datasets considered here, a single variational inference run can yield posterior samples of the physical parameters rather than a single point estimate. 
This computational efficiency makes it practical to compare multiple physically motivated ansatzes and investigate competing explanations of experimentally generated quantum states.
Regardless of the specific inference strategy employed, the proposed framework provides physicists with a probabilistic characterization for identifying the physical mechanisms underlying experimentally generated mixed quantum states.

\section*{Acknowledgments}
This work is partially supported by the National Science and Technology Council of Taiwan (Nos
112-2123-M-007-001, 112-2119-M-008-007, 114-2112-M-007-044-MY3, 115-2112-M-007-026-MY3), Office of Naval Research Global, and the collaborative research program of the Institute for Cosmic Ray Research (ICRR) at the University of Tokyo.


\bibliography{ref}

\clearpage
\raggedbottom
\setcounter{section}{0}
\renewcommand{\thesection}{SI~\arabic{section}}
\renewcommand{\thefigure}{SI.~\arabic{figure}}
\setcounter{figure}{0}
\setcounter{equation}{0}
\renewcommand{\theequation}{SI.~\arabic{equation}} 
\renewcommand{\thetable}{SI.~\arabic{table}}
\renewcommand*{\theHtable}{\thetable}
\renewcommand*{\theHfigure}{\thefigure}
\renewcommand*{\theHsection}{\thesection}
\renewcommand*{\theHequation}{\theequation}

\onecolumngrid

\begin{center}
  {\large \bf -- Supplementary Information -- \\
  \vspace{0.25cm}
  Normalizing Flow-Based Bayesian Parameter Estimation for Noisy Quantum States}
  \\
  \vspace{0.25cm}

  {{Hsien-Yi Hsieh\orcidlink{0000-0001-5227-8248}}, 
  {Yuan-Ting Liu}, 
  {Juan Camilo Rodr\'iguez\orcidlink{0009-0005-2128-3016}}, 
  {Yang-Yi Lee}, 
  {Po-Hang Wang\,\orcidlink{0009-0009-3800-5433}}, 
  {Ole Steuernagel\orcidlink{0000-0001-6089-7022}}, 
  {Chien-Ming Wu}, 
  and Ray-Kuang Lee\orcidlink{0000-0002-7171-7274}}
\end{center}

\onecolumngrid
\setcounter{figure}{0}

\section{Analysis of Dataset 2}
\label{sec:dataset2}

\begin{figure}[H]
    \centering
    \includegraphics[width=10cm]{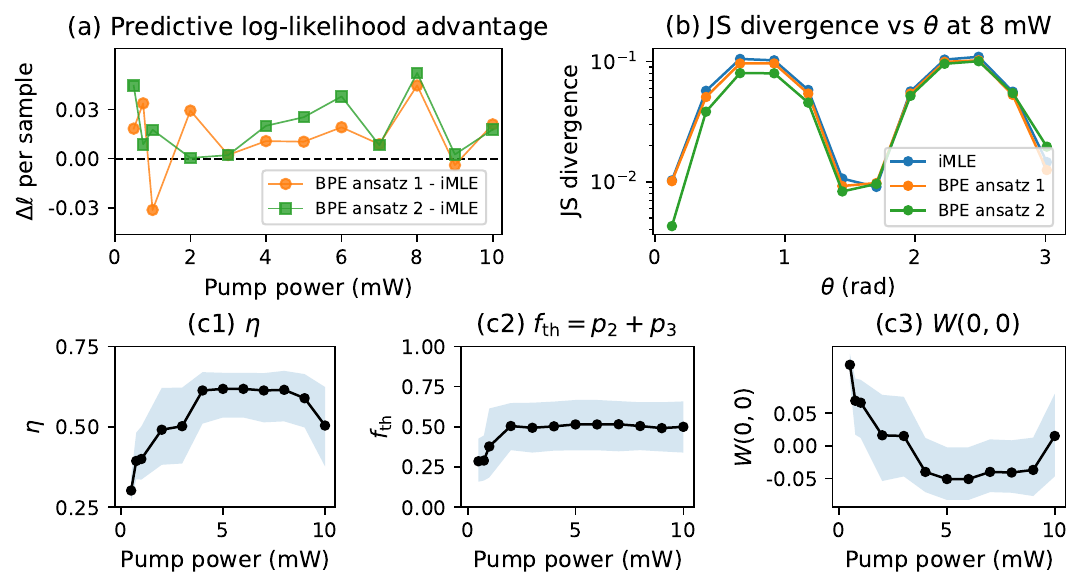}
    \caption{Predictive performance and inferred physical parameters for
    {\it Dataset 2}.
    (a) Average log-likelihood difference,
    $\Delta\ell=\ell_{\rm BPE}-\ell_{\rm iMLE}$, per quadrature sample.
    Positive values indicate improved agreement relative to the iMLE
    reconstruction.
    (b) Phase-resolved Jensen--Shannon (JS) divergence between the measured
    and predicted quadrature distributions for the 8 mW StPDC setting.
    Posterior medians of
    (c1) the photon-addition fraction $\eta$,
    (c2) the thermal fraction $f_{\rm th}=p_2+p_3$, and
    (c3) the Wigner function at the origin $W(0,0)$
    as functions of StPDC pump power.
    Shaded regions denote $68\%$ credible intervals.}
    \label{fig:main_spdc}
\end{figure}

Here we provide the complementary analysis of {\it Dataset 2}, in which
the StPDC pump power is varied while the OPO pump power is fixed.
Fig.~\ref{fig:main_spdc} summarizes the predictive performance and
inferred physical parameters using the same analysis procedure as for
{\it Dataset 1} in the main text.

As shown in Fig.~\ref{fig:main_spdc}(a), at least one of the two candidate
ansatzes provides a likelihood comparable to or greater than that of the
iMLE reconstruction throughout the investigated StPDC pump-power range.
Taking the 8 mW dataset as a representative example,
Fig.~\ref{fig:main_spdc}(b) provides a further comparison using the
phase-resolved JS divergence. The three estimates exhibit similar
phase-dependent trends. Near $\theta=0$, $\mathcal A_2$ yields a smaller
JS divergence and thus better agreement with the data, whereas the
opposite tendency is observed near $\theta=\pi$. When averaged over
$\theta$, $\mathcal A_2$ yields a slightly smaller JS divergence than
$\mathcal A_1$. Nevertheless, the differences between the two ansatzes
are small in both the average JS divergence and the average
log-likelihood comparison shown in Fig.~\ref{fig:main_spdc}(a).

Figs.~\ref{fig:main_spdc}(c1)--(c3) show the corresponding evolution of
the inferred physical quantities. As the StPDC pump power increases,
the photon-addition fraction $\eta$ initially increases and subsequently
approaches a plateau near $\eta\sim0.6$. The thermal fraction
$f_{\rm th}=p_2+p_3$ likewise increases before remaining approximately
constant over much of the investigated range. In contrast to the more
systematic evolution observed when varying the OPO pump power in
{\it Dataset 1}, $W(0,0)$ exhibits a nonmonotonic dependence on the StPDC
pump power: the Wigner negativity initially increases in magnitude and
then remains broadly stable over the intermediate-power range before
weakening again at the highest power.

\section{Constraint mapping}
\label{sec:constraint mapping}
The normalizing flow produces unconstrained latent variables
$\mathbf{u}\in\mathbb{R}^{d}$.
To ensure that the inferred parameters satisfy the physical constraints of the quantum-state model, we introduce a deterministic constraint mapping
\begin{equation}
\mathbf{s}=g(\mathbf{u}),
\end{equation}
which maps the unconstrained outputs of the flow onto the physically admissible parameter space.

The mapping consists of two components acting on disjoint subsets of the flow outputs,
\begin{equation}
g
=
g_{\mathrm{simplex}}
\oplus
g_{\mathrm{bounded}},
\end{equation}
corresponding to simplex-constrained parameters and bounded physical parameters, respectively.

\paragraph*{Bounded parameters.}

Parameters constrained to finite intervals
\[
s_i\in[l_i,h_i]
\]
are obtained using a sigmoid transformation,
\begin{equation}
y_i=\sigma(u_i),
\qquad
s_i=l_i+(h_i-l_i)y_i.
\end{equation}

The corresponding log-determinant of the Jacobian is
\begin{equation}
\log\left|\det J_{\mathrm{bounded}}\right|
=
\sum_i
\log(h_i-l_i)
+
\sum_i
\log\!\left(
y_i(1-y_i)
\right).
\end{equation}

\paragraph*{Simplex parameters.}
The mixture weights
$(p_1,p_2,p_3)$
are mapped onto the simplex using a stick-breaking transformation~\cite{Blei2006}.

First define
\begin{equation}
v_k=\sigma(u_k),
\qquad
k=1,\ldots,m-1.
\end{equation}

The simplex coordinates are constructed sequentially as
\begin{align}
p_1 &= v_1, \\
p_2 &= (1-v_1)v_2, \\
p_3 &= (1-v_1)(1-v_2).
\end{align}

Using the first $m-1$ simplex coordinates as independent variables, the reduced Jacobian is lower triangular. Defining
\begin{equation}
\mathrm{stick}_{k-1}
=
\prod_{j<k}(1-v_j),
\end{equation}
the corresponding log-determinant is
\begin{equation}
\log\left|\det J_{\mathrm{simplex}}\right|
=
\sum_{k=1}^{m-1}
\left[
\log(\mathrm{stick}_{k-1})
+
\log\!\left(
v_k(1-v_k)
\right)
\right].
\end{equation}

\paragraph*{Combined transformation.}

The complete constraint mapping is therefore
\begin{equation}
g
=
g_{\mathrm{simplex}}
\oplus
g_{\mathrm{bounded}},
\end{equation}
whose Jacobian factorizes as
\begin{equation}
\log\left|\det J_g\right|
=
\log\left|\det J_{\mathrm{simplex}}\right|
+
\log\left|\det J_{\mathrm{bounded}}\right|.
\end{equation}

Since both transformations admit closed-form Jacobians, no numerical approximation of the Jacobian determinant is required. Instead, the analytical Jacobian expressions are implemented directly as differentiable PyTorch~\cite{PyTorch} modules.

\section{Prior calibration}
\label{sec:prior calibration}

As introduced in the main text, the calibrated prior
$p(\boldsymbol{s}\mid D_C,\epsilon)$ combines empirical prior information
$\epsilon$ with the calibration subset $D_C$. The empirical information
$\epsilon$ specifies physically plausible parameter ranges and a finite
set of candidate parameter configurations, whereas $D_C$ is used only to
select a coarse representative configuration from these candidates.
The selected configuration then determines the means of the resulting
weakly informative priors.

\paragraph*{Prior distributions and parameterization.}
For a bounded physical parameter $s_j\in[l_j,h_j]$, we introduce the
normalized variable
\begin{equation}
y_j=\frac{s_j-l_j}{h_j-l_j}\in[0,1],
\end{equation}
and assign the Beta prior
\begin{equation}
y_j\sim\operatorname{Beta}(\alpha_j,\beta_j),
\end{equation}
whose probability density is denoted by
$\operatorname{Beta}(y_j;\alpha_j,\beta_j)$.
Rather than working directly with the conventional shape parameters
$(\alpha_j,\beta_j)$, we reparameterize the distribution in terms of
a normalized prior center $\mu_j\in(0,1)$ and a concentration parameter
$m_j>0$~\cite{reparametrization1,reparametrization2},
\begin{equation}
\alpha_j=m_j \mu_j,
\qquad
\beta_j=m_j(1-\mu_j).
\label{eq:beta_reparam}
\end{equation}
Here, $\mu_j$ controls the prior mean, whereas $m_j$ controls its
concentration, or strength. In particular,
\begin{equation}
\mathbb{E}[y_j]=\mu_j,
\qquad
\mathbb{E}[s_j]
=
l_j+(h_j-l_j)\mu_j.
\label{eq:beta_mean}
\end{equation}

To construct weak, log-concave priors while avoiding boundary-singular
densities, we require
\begin{equation}
\label{eq:concentraion_1}
m_j\mu_j\geq1,
\qquad
m_j(1-\mu_j)\geq1,
\end{equation}
and choose $m_j$ near the smallest integer value satisfying these
conditions. This avoids strongly boundary-dominated priors while keeping
their concentration low.

For probability vectors
$\boldsymbol{\xi}=(\xi_1,\ldots,\xi_K)$ on the simplex,
$\xi_k\geq0$ and $\sum_k\xi_k=1$, we analogously employ a Dirichlet prior,
\begin{equation}
\boldsymbol{\xi}
\sim
\operatorname{Dir}(\boldsymbol{\alpha}),
\qquad
\alpha_k=m \mu_k,
\qquad
\sum_k \mu_k=1,
\label{eq:dirichlet_reparam}
\end{equation}
with probability density denoted by
$\operatorname{Dir}(\boldsymbol{\xi};\boldsymbol{\alpha})$.
The vector $\boldsymbol{\mu}$ specifies the prior mean,
\begin{equation}
\mathbb{E}[\xi_k]=\mu_k,
\end{equation}
while $m$ controls the overall concentration. We require
\begin{equation}
m \mu_k\geq1
\qquad
\text{for all }k,
\label{eq:concentraion_2}
\end{equation}
and again choose $m$ near the smallest integer value satisfying this
condition.
\medskip
\paragraph*{Coarse calibration from homodyne data.}
We next describe how $D_C$ is used to select a coarse representative
parameter configuration from the finite candidate set specified by
$\epsilon$. The selected configuration is used only to determine the
means of the weakly informative priors and is not itself interpreted as an estimator of the quantum state.

For a physical parameter configuration $\boldsymbol{s}$, we discretize
the phase interval $\theta\in[0,\pi]$ and the quadrature interval
$x\in[-5,5]$ into 64 bins each. The predicted probability of a homodyne
event in the $(k,l)$th bin defines the elements of a $64\times64$ matrix,
\begin{equation}
M_{kl}(\boldsymbol{s})
=
\operatorname{Tr}
\!\left[
\Pi(\theta_k,x_l)\rho(\boldsymbol{s})
\right]\Delta x ,
\label{eq:prior_summary_matrix}
\end{equation}
which provides a coarse summary of the predicted homodyne statistics.
An empirical counterpart $\hat{M}$ is constructed from $D_C$ by
two-dimensional histogram estimation. For each candidate parameter
configuration, we compare the simulated and empirical summary statistics
using
\begin{equation}
d(\boldsymbol{s})
=
\sum_{k,l}
\left[
\hat{M}_{kl}-M_{kl}(\boldsymbol{s})
\right]^2 .
\label{eq:prior_distance}
\end{equation}

The calibration is performed in two stages. First, the squeezing angle
$\phi$ is treated separately because it produces a pronounced geometric
variation in the phase-resolved homodyne statistics. The remaining
parameters are fixed at representative values,
\begin{equation}
p_1=p_2=p_3=\frac{1}{3},
\quad
\eta=0.5,
\quad
r=0.5,
\quad
\bar n_1=\bar n_2=0.15,
\end{equation}
while $\phi$ is scanned over $[0,2\pi]$. We select
\begin{equation}
\phi^\star
=
\underset{\phi}{\arg\min}\;
d(\phi),
\label{eq:phi_prior_selection}
\end{equation}
and keep $\phi^\star$ fixed in the subsequent coarse scan.

In the second stage, the remaining parameters are restricted to a finite
set of candidate values informed by the empirical prior information
$\epsilon$. For the simplex parameters, we consider
\begin{equation}
\begin{aligned}
(p_1,p_2,p_3)\in\{&
(0.8,0.1,0.1),(0.7,0.2,0.1),
(0.5,0.3,0.2),\\
&(0.5,0.4,0.1),(0.6,0.3,0.1),
(0.6,0.2,0.2)\},
\end{aligned}
\label{eq:prior_candidates_simplex}
\end{equation}
together with
\begin{equation}
\begin{aligned}
\eta &\in \{0.5,0.6,0.7\},\\
r &\in \{0.2,0.3,0.4,0.5\},\\
\bar n_1,\bar n_2 &\in \{0.1,0.2,0.3\}.
\end{aligned}
\label{eq:prior_candidates_bounded}
\end{equation}
These candidate sets encode the empirical prior information $\epsilon$;
for example, the mixture candidates reflect the expectation that the
squeezed-state component is dominant in the experimental regime
considered here. Additional parameter-specific information can likewise
be incorporated through the admissible ranges; in particular, the
squeezing range is informed by the independent spectral-analyzer
measurements described in Sec.~\ref{sec:sa_measurements}.

With $\phi=\phi^\star$ fixed, a representative configuration is selected as
\begin{equation}
\boldsymbol{s}^{\star}
=
\underset{\boldsymbol{s}\in\mathcal{C}_{\epsilon}}
{\arg\min}\;
d(\boldsymbol{s}),
\label{eq:coarse_prior_center}
\end{equation}
where $\mathcal{C}_{\epsilon}$ denotes the finite candidate set specified
above, with $\phi$ fixed at $\phi^\star$. This yields
\begin{equation}
\boldsymbol{s}^{\star}
=
(p_1^\star,p_2^\star,p_3^\star,
\eta^\star,r^\star,\phi^\star,
\bar n_1^\star,\bar n_2^\star).
\end{equation}
We emphasize that $\boldsymbol{s}^{\star}$ is a heuristic anchor for
prior calibration rather than a point estimator of the underlying
quantum state.

\paragraph*{Construction of the calibrated prior.}
The selected configuration $\boldsymbol{s}^{\star}$ determines the means of the weakly informative priors. For the simplex parameters,
\begin{equation}
\boldsymbol{\mu}
=
(p_1^\star,p_2^\star,p_3^\star),
\end{equation}
which specifies the mean of the corresponding Dirichlet distribution.

For each bounded parameter $s_j^\star\in[l_j,h_j]$, the corresponding
normalized prior mean is
\begin{equation}
\mu_j
=
\frac{s_j^\star-l_j}{h_j-l_j},
\label{eq:prior_mean_mapping}
\end{equation}
such that Eq.~(\ref{eq:beta_mean}) gives
$\mathbb{E}[s_j]=s_j^\star$

The concentration parameters are then chosen according to
Eqs.~(\ref{eq:concentraion_1}) and
(\ref{eq:concentraion_2}), yielding weakly informative priors whose means correspond to the selected coarse configuration. Because $\phi$ has a particularly pronounced geometric signature in the
phase-resolved homodyne statistics, we use a moderately more concentrated
prior for this parameter, with $m_\phi$ chosen to be five times the
minimum value satisfying Eq.~(\ref{eq:concentraion_1}).

For the photon-addition model considered in the main text,
$\boldsymbol{s}_{\mathrm{simplex}}=(p_1,p_2,p_3)$ and
$\boldsymbol{s}_{\mathrm{bounded}}
=(\eta,r,\phi,\bar n_1,\bar n_2)$.
Assuming independent priors for the bounded parameters, the resulting
calibrated prior is therefore
\begin{equation}
\begin{aligned}
p(\boldsymbol{s}\mid D_C,\epsilon)
={}&
\operatorname{Dir}
\!\left(
\boldsymbol{s}_{\mathrm{simplex}};
m\boldsymbol{\mu}
\right)
\\
&\times
\prod_j
\frac{1}{h_j-l_j}
\operatorname{Beta}
\!\left(
\frac{s_{\mathrm{bounded}}^j-l_j}{h_j-l_j};
m_j\mu_j,m_j(1-\mu_j)
\right).
\end{aligned}
\label{eq:calibrated_prior}
\end{equation}

\section{Squeezing and anti-squeezing measurements}
\label{sec:sa_measurements}

Three additional spectral-analyzer measurements were performed at OPO
pump powers of 20, 30, and 40 mW to characterize the squeezing and
anti-squeezing levels, as summarized in Table~\ref{tab:sa_measurements}.
The measured anti-squeezing (ASQ) level is used to determine the upper
bound of the prior support for the squeezing parameter $r$. Specifically,
the measured value $r_{\rm dB}$ is converted to the squeezing parameter
according to
\begin{equation}
r=\frac{r_{\rm dB}}{20}\ln 10.
\label{eq:r_db_conversion}
\end{equation}
The lower bound is fixed at $r=0$, such that the resulting prior support
is $r\in[0,r_{\max}]$, with $r_{\max}$ determined from the corresponding
ASQ measurement. We emphasize that the converted value is used only as an upper bound of the prior support; $r$ itself is inferred from $D_I$ through the likelihood. Each auxiliary measurement is used to set the prior support for the
homodyne data with the corresponding OPO setting, as indicated in
Table~\ref{tab:sa_measurements}; since the OPO power is fixed at 20 mW
in {\it Dataset 2}, the corresponding support is used throughout that
dataset.

\begin{table}[h]
\centering
\caption{Auxiliary squeezing/anti-squeezing measurements and the
corresponding prior supports for the squeezing parameter $r$.}
\begin{tabular}{ccccc}
\toprule
OPO pump power
& ASQ (dB)
& SQ (dB)
& Prior support of $r$
& Applied homodyne data \\
\midrule
20 & 5.33 & 4.01 & $[0,0.614]$
& Dataset 1: 15, 20 mW; Dataset 2 \\
30 & 6.04 & 4.50 & $[0,0.695]$
& Dataset 1: 25, 30 mW \\
40 & 6.27 & 4.57 & $[0,0.722]$
& Dataset 1: 35, 40 mW \\
\bottomrule
\end{tabular}
\label{tab:sa_measurements}
\end{table}
\section{Homodyne measurement datasets}

\begin{table}[h]
\centering
\caption{Dataset 1: fixed StPDC pump power and varied OPO pump power.}
\begin{tabular}{cccc}
\toprule
 StPDC pump power (mW) & OPO pump power (mW) & Number of quadrature points \\
\midrule
10 & 15 & 69680 \\
10 & 20 & 73943 \\
10 & 25 & 69680 \\
10 & 30 & 79717 \\
10 & 35 & 71106 \\
10 & 40 & 56508 \\
\bottomrule
\end{tabular}
\label{tab:dataset1}
\end{table}

\begin{table}[h]
\centering
\caption{Dataset 2: fixed OPO pump power and varied StPDC pump power.}
\begin{tabular}{cccc}
\toprule
OPO pump power (mW) & StPDC pump power (mW) & Number of quadrature points \\
\midrule
20 & 0.25 & 13976 \\
20 & 0.50 & 16287 \\
20 & 0.75  & 16240 \\
20 & 1.00 & 30154 \\
20 & 2.00 & 29844 \\
20 & 3.00 & 43470 \\
20 & 4.00 & 33344 \\
20 & 5.00 & 39364 \\
20 & 6.00 & 56398 \\
20 & 7.00 & 42444 \\
20 & 8.00 & 70033 \\
20 & 9.00 & 77208 \\
20 & 10.00 & 61863 \\
\bottomrule
\end{tabular}
\label{tab:dataset2}
\end{table}

\section{Parameters of physical ansatzes}

The parameters of the physical ansatzes are summarized here.
Since ansatz \(\mathcal{A}_2\) extends ansatz \(\mathcal{A}_1\),
only the additional parameters introduced in \(\mathcal{A}_2\)
are listed separately.

\begin{table}[H]
\centering
\caption{Parameters appearing in ansatz $\mathcal{A}_1$.}
\begin{tabular}{lll}
\toprule
Parameter & Domain & Physical meaning \\
\midrule
$p_1,p_2,p_3$ & Simplex & Mixture weights of the degraded component \\
$\eta$ & $[0,1]$ & Photon-added fraction \\
$r$ & Bounded & Squeezing magnitude \\
$\phi$ & Bounded / periodic & Squeezing angle \\
$\bar n_1$ & Bounded & Thermal occupation in squeezed thermal component \\
$\bar n_2$ & Bounded & Thermal occupation in thermal component \\
\bottomrule
\end{tabular}
\label{tab:a1_params}
\end{table}

\begin{table}[H]
\centering
\caption{Additional parameters appearing in ansatz $\mathcal{A}_2$.}
\begin{tabular}{lll}
\toprule
Parameter & Domain & Physical meaning \\
\midrule
$|\beta_0|$ & Bounded & Displacement magnitude of degraded component \\
$|\beta_1|$ & Bounded & Displacement magnitude of photon-added component \\
$\arg(\beta_0)$ & Bounded / periodic & Displacement phase of degraded component \\
$\arg(\beta_1)$ & Bounded / periodic & Displacement phase of photon-added component \\
\bottomrule
\end{tabular}
\label{tab:a2_params}
\end{table}

\section{ Pseudo code}
\noindent
\textbf{Stage 1: Data Splitting}

\begin{algorithmic}[1]
    \State Divide the measurement data $D=\left\{x_i,\theta_i \right\}_{i=1}^{N}$ into inference $D_I$(80\%) and calibration $D_C$(20\%) sets.
    \State Divide the inference data into training (80\%) and validation (20\%) sets.
\end{algorithmic}

\vspace{0.5em}
\noindent
\textbf{Stage 2: Prior function construction}

\begin{algorithmic}[1]
    \State Apply the method described in Sec.~\ref{sec:prior calibration} to the calibration data $D_C$ to construct the prior $p(\boldsymbol{s}\mid D_C,\epsilon )$, where \(\epsilon\) collectively denotes empirical prior information, including physically motivated parameter ranges and auxiliary calibration knowledge used in constructing the prior, as discussed in in Sec.~\ref{sec:prior calibration} and Sec.~\ref{sec:sa_measurements}.
    \State Evaluate $\log p(\boldsymbol{s}| D_C,\epsilon).
= \log p(\boldsymbol{s}_{\mathrm{simplex}}| D_C,\epsilon)+\sum_j \log p(s_{\mathrm{bounded}}^j| D_C,\epsilon)$
\end{algorithmic}

\vspace{0.5em}

\noindent
\textbf{Stage 3: Likelihood function construction}

\begin{algorithmic}[1]
    \State Construct the functions $\rho(\boldsymbol{s})$ for the chosen physical ansatz.
    \State Construct the function $ 
\log p(D_I\mid\mathbf{s})
=
\sum_{(x_i,\theta_i)\in D_I}\log
\operatorname{Tr}
\!\left[
\Pi(x_i,\theta_i)\rho(\mathbf{s})
\right]$.
\end{algorithmic}

\vspace{0.5em}






\vspace{0.5em}
\noindent
\textbf{Stage 4: Training loop and inference}
\begin{algorithmic}[1]
    \State Initialize the normalizing-flow parameters $\mathbf{w}$.
    \State Set the early-stopping counter to zero.
\end{algorithmic}

\begin{algorithmic}
    \For{$e=1,\ldots,N_{\mathrm{epoch}}$}

        \State Draw latent samples
        $\mathbf{z}\sim\mathcal{N}(\mathbf{0},\mathbf{I})$

        \State Compute
        $\mathbf{s}
        =
        T_{\mathbf{w}}(\mathbf{z})$ and Jacobians $J_f$ and $J_g$

        \State Evaluate $\mathcal{L}(\mathbf{w})$

        \State Update $\mathbf{w}$ using Adam

        \If{the validation loss does not improve}
            \State Increase the early-stopping counter
        \Else
            \State Save the current model
            \State Reset the early-stopping counter
        \EndIf

        \If{the early-stopping criterion is satisfied}
            \State \textbf{break}
        \EndIf

    \EndFor
\end{algorithmic}

\vspace{0.5em}
\noindent
\textbf{Stage 5: Inference}
\begin{algorithmic}[1]
    \State Draw posterior samples
    $\mathbf{s}=T_{\mathbf{w}}(\mathbf{z})\sim q_{\mathbf{w}}(\mathbf{s})$.
    \State Compute posterior summaries and credible intervals.
\end{algorithmic}
\section{Notation}

\begin{table}[h]
\centering
\caption{Summary of notation used in the main text and Supplementary Material.}
\begin{tabular}{lll}
\toprule
Symbol & Meaning & Role (played in this work) \\
\midrule
$D$ & Homodyne dataset $\{x_i,\theta_i\}_{i=1}^{N}$ & Experimental dataset \\
$D_C$ & Calibration dataset  & Used for prior calibration \\
$D_I$ & Inference dataset  & Used for likelihood evaluation \\
$\epsilon$ & Empirical knowledge  & Empirical knowledge \\
$x_i$ & Quadrature outcome & Data point \\
$\theta_i$ & Local-oscillator phase & LO setting \\
$\boldsymbol{s}$ & Physical model parameters & Inference target \\
$\mathbf{w}$ & Neural-network parameters & Trainable Variational Inference parameters \\
$\mathbf{z}$ & Flow Input (base latent) variable & Sampled from $p_z(\mathbf{z})$ \\
$\boldsymbol{u}$ & Flow output variable & Here: Input to constraint mapping $g$ \\
$f$ & Normalizing flow mapping & Maps $\mathbf{z}$ to $\boldsymbol{u}$ \\
$g$ & Constraint mapping & Maps $\boldsymbol{u}$ to physical parameters $\boldsymbol{s}$ \\
$T=g\circ f$ & Full transformation & Maps $\mathbf{z}$ to $\boldsymbol{s}$ \\
$q_s(\boldsymbol{s};\mathbf{w})$ & Variational posterior & Approximation to $p(\boldsymbol{s}| D_I,D_C,\epsilon)$ \\
$p(D|\boldsymbol{s})$ & Data likelihood & Generic Bayesian likelihood \\
$p(\boldsymbol{s})$ & Prior distribution & Generic Bayesian prior \\
$p(\boldsymbol{s}|D)$ & Posterior distribution & Generic Bayesian posterior\\
$p(\boldsymbol{s}| D_C,\epsilon )
$ & Calibrated prior & Based on $D_C$ \\
$p(D_I|\boldsymbol{s})$
& Inference-data likelihood
& Likelihood evaluated on $D_I$ \\
$\rho(\boldsymbol{s})$ & Parameterized density matrix & Ansatz for physical state \\
$\eta$ & Photon-added fraction & A physical parameter \\
$p_1,p_2,p_3$ & Mixture weights & Simplex-constrained parameters \\
$r,\phi$ & Squeezing magnitude and angle & Physical parameters \\
$\bar n_1,\bar n_2$ & Thermal photon numbers & Physical parameters \\
$\beta_0,\beta_1$ & Displacement parameters & Parameters for back-scattering ansatz  \\
$\mathcal{A}_1,\mathcal{A}_2$ & Candidate physical ansatzes & Models compared in experimental analysis \\
\bottomrule
\end{tabular}
\label{tab:notation}
\end{table}

\clearpage
\end{document}